\documentclass[sigconf, language=english]{acmart}

\usepackage{booktabs}
\usepackage{subfig}
\usepackage{xcolor}
\usepackage{enumitem}

\AtBeginDocument{%
  \providecommand\BibTeX{{%
    \normalfont B\kern-0.5em{\scshape i\kern-0.25em b}\kern-0.8em\TeX}}}

\copyrightyear{2023}
\acmYear{2023}
\setcopyright{rightsretained}
\acmConference[WWW '23]{Proceedings of the ACM Web Conference 2023}{May 1--5, 2023}{Austin, TX, USA}
\acmBooktitle{Proceedings of the ACM Web Conference 2023 (WWW '23), May 1--5, 2023, Austin, TX, USA}
\acmDOI{10.1145/3543507.3583514}
\acmISBN{978-1-4503-9416-1/23/04}

\begin{document}

\title[Sedition Hunters]{Sedition Hunters: A Quantitative Study of the Crowdsourced Investigation into the 2021 U.S. Capitol Attack}

\author{Tianjiao Yu}
\affiliation{%
  \institution{Virginia Tech}
  \city{Blacksburg}
  \state{VA}
  \country{USA}
}
\email{tianjiao@vt.edu }

\author{Sukrit Venkatagiri}\authornote{A portion of this work was completed while this author was at Virginia Tech.}
\affiliation{%
  \institution{University of Washington}
  \city{Seattle}
  \state{WA}
  \country{USA}
}
\email{sukritv@uw.edu}

\author{Ismini Lourentzou}
\affiliation{%
  \institution{Virginia Tech}
  \city{Blacksburg}
  \state{VA}
  \country{USA}
}
\email{ilourentzou@vt.edu}

\author{Kurt Luther}
\affiliation{%
  \institution{Virginia Tech}
  \city{Arlington}
  \state{VA}
  \country{USA}
}
\email{kluther@vt.edu}

\renewcommand{\shortauthors}{Yu et al.}

\begin{abstract}
Social media platforms have enabled extremists to organize violent events, such as the 2021 U.S. Capitol Attack. Simultaneously, these platforms enable professional investigators and amateur sleuths to collaboratively collect and identify imagery of suspects with the goal of holding them accountable for their actions. Through a case study of Sedition Hunters, a Twitter community whose goal is to identify individuals who participated in the 2021 U.S. Capitol Attack, we explore what are the main topics or targets of the community, who participates in the community, and how. Using topic modeling, we find that information sharing is the main focus of the community. We also note an increase in awareness of privacy concerns. Furthermore, using social network analysis, we show how some participants played important roles in the community. Finally, we discuss implications for the content and structure of online crowdsourced investigations.
\end{abstract}



\maketitle

\section{Introduction}

On January 6, 2021, the U.S. Capitol was violently attacked by a group of more than 2,000 Donald Trump supporters aiming to nullify the formalization of Joe Biden’s presidential victory. This was the most severe attack on the U.S. Capitol since the War of 1812, causing several deaths, hundreds of injuries, and at least \$30 million in property damage and related security expenses \cite{cochrane_capitol_2021, chappell2021architect}. It is also arguably ``the most documented crime in U.S. history'' \cite{Bergengruen2021theCapitol} as there is evidence collected not only from surveillance infrastructure but also social media where hundreds of rioters were live streaming and uploading posts during and after the event \cite{bergengruen_capitol_2021}. Following the Capitol Attack, the U.S. Department of Justice (DOJ) and Federal Bureau of Investigation (FBI) began investigations that have led to over 840 arrests and 185 criminal sentences (as of June 2022) \cite{whatHappenedJan6}. However, hundreds of suspects remain unidentified due to the large scale of the event, prompting law enforcement officials from the FBI and the Washington, D.C. police department to seek help from the public \cite{FBI_mostwanted}.

Among the members of the public assisting in suspect identification, a Twitter community quickly formed around the hashtag \texttt{\#SeditionHunters}. A related account, \texttt{@SeditionHunters}, was created on January 12, 2020, and gathered more than 66,000 followers. This movement of online sleuthing or web-sleuthing \cite{yardley2018s, nhan2017digilantism} harnessed the power of collective knowledge and resources to assist the police in their criminal investigation. They described themselves as Open-Source Intelligence (OSINT) investigators working together to identify suspects who allegedly committed crimes in the January 6 Capitol Attack \cite{seditionhunterWeb}. The Sedition Hunters community is notable for their significant contributions to multiple successful arrests \cite{ mak2021fbiUsingClues} and legal proceedings \cite{legal-proceeding-1, legal-proceeding-2}. Over a year later, the \texttt{\#SeditionHunters} community remains active, in contrast to most similar gatherings \cite{smallridge2016understanding}.

Much discussion has been made about the potential value of crowdsourced data contributions to law enforcement \cite{palen2008emergence, rizza2014yourself}. The investigation of the Vancouver Hockey Riots of 2011 is an early example illustrating how the Internet offers a venue for the public and law enforcement officials to achieve a shared objective through pooled intelligence and resources \cite{schneider2013social}. Despite the growing value of crowdsourced investigations, they have also raised serious concerns. For example, as police actively incorporated the public's help to provide information about the suspects who planted bombs at the 2013 Boston Marathon, the crowd yearned to be more than information suppliers. They sought to participate in the crime-solving process, which ultimately led to four misidentifications and harmful consequences for the victims \cite{nhan2017digilantism}. 

Analyzing patterns of public participation in online cyber-policing gives insights into when and how collaborative efforts can bring about positive or negative impacts during the so-called Information Era of policing \cite{kelling1989evolving}. It also provides insights into how crowdsourced investigations can be better harnessed for efficiency and effectiveness. Given the most researched large-scale crowdsourced criminal investigation event, the 2013 Boston Marathon Bombing, happened almost a decade ago, Sedition Hunters provide us a unique opportunity to investigate how such a community deals with the ever-growing overabundance of data \cite{capitolriotsize}, how recent technology (e.g., Face recognition) can be utilized for better output, and how the community is organized and regulated for improved effectiveness.

In this paper, we use quantitative approaches, namely topic modeling and social network analysis, to conduct an analysis of more than 300,000 posts and 65,282 unique users related to the Sedition Hunters community, spanning exactly one year from the Capitol Attack. We find that the content of most posts focused on information sharing, which included details about suspects and related news and updates. Topic analysis also reveals the tracking tools the community used for the investigation process such as hashtags, spreadsheets, and self-developed websites. 
Social network analysis shows that, as most participants did not engage comprehensively, a few dominant accounts are responsible for multiple roles of gathering, processing, and distributing information. We conclude by discussing the factors that are unique to the Sedition Hunters community and contributed to its success.

\section{Related Work}

With the prevalence of portable computing devices and the desire to document daily activities, modern society empowers people to provide huge amounts of persistent, searchable, and remotely accessible archives of formerly private information and events \cite{marwick2012public}. The openness and interactivity of social media also encourage crowdsourced investigations, which use the knowledge, work, and content of online communities as a vastly extended surveillance network \cite{walsh2019social}. Simultaneously, the entire format of policing has advanced into the Information Era in which the crime-fighting process has become data-driven, intelligence-led, and technologically mediated \cite{kelling1989evolving}. Law enforcement officials are considered more as ``knowledge workers'' who now have more responsibility for collecting and processing information \cite{ericson1997policing}. Our analysis of Sedition Hunters highlights the complex and nuanced role the public sometimes seeks to play in online crime investigation. Further, we see that the Sedition Hunters community is able to self-organize regulatory and other strategies by which they can direct themselves in ways that generate useful information more efficiently while minimizing negative impacts such as misidentification. 

Previous studies posited three primary investigative models for solving crimes. Trottier \cite{daniel2014police} studied how traditional top-down policing by investigative agencies and bottom-up policing by crowdsourced users converge and co-exist on social media. \cite{venkatagiri2021crowdsolve} studied expert-led crowdsourced investigations, a hybrid of top-down and bottom-up criminal investigation models, in the hope of balancing the tension between experts and crowds to achieve greater results. Sedition Hunters is different from previous investigative models because, while most participants of the Sedition Hunters are amateurs, some members claim they have expertise or previous experience in OSINT \cite{reilly2021sedition}. In other words, the experts are part of the crowd. The Sedition Hunters community was able to self-regulate the interactions between experienced members and novices, in which they created relatively standardized procedures for the investigation, such as creating composite suspect images, releasing information with the \#doyouknow hashtag, and building progress-tracking and coordination websites.

Scholars use terms such as digilantism \cite{nhan2017digilantism} and web-sleuthing \cite{yardley2018s} to describe cases where members of the public take part in online investigations. Tapia and Lalone used sentiment analysis to examine moments during crowdsourced crisis investigation when the public became either more positively or negatively inclined toward the acts of the web-sleuthing participants, reflecting social approval or disapproval of such actions \cite{tapia2014crowdsourcing}. Marx also qualitatively examined opportunities and risks of crowdsourced efforts and implications for issues of justice given the new technologies \cite{marx2013public}. Compared with these studies, we use quantitative methods to explore the content and participants of this specific online crowdsourced community, the Sedition Hunters.

Partially due to the limited scale of most crowdsourced crime investigations and their dispersed nature, prior work has focused on how they play a role in criminology in a broader sense. Sedition Hunters provided an opportunity to study the community itself and analyze how such a large-scale crowdsourced investigation community successfully delivers useful information. Most related to our work is the study of Venkatagiri et al. who conducted a mixed-methods analysis of the Sedition Hunters community. They presented preliminary results regarding who are the sedition hunters and what did they do to produce successful results \cite{venkatagiri2021sedition}. Our research extends this study in an in-depth, quantitative approach with topic modeling and social network analysis, along with a focused manual qualitative examination to further investigate how and why the Sedition Hunters do what they do to produce successful results. This paper addresses the following research questions:
\begin{itemize}
    \item \textbf{RQ1}: What are the discussions within and around the Sedition Hunters community? What are the main topics or targets of \#SH?
    \item \textbf{RQ2}: Who participates in the Sedition Hunters community and how? What are the different roles in the SH community? What methods have they used?
\end{itemize}

\begin{figure*}[t!]
    \centering
    \includegraphics[width=0.8\linewidth]{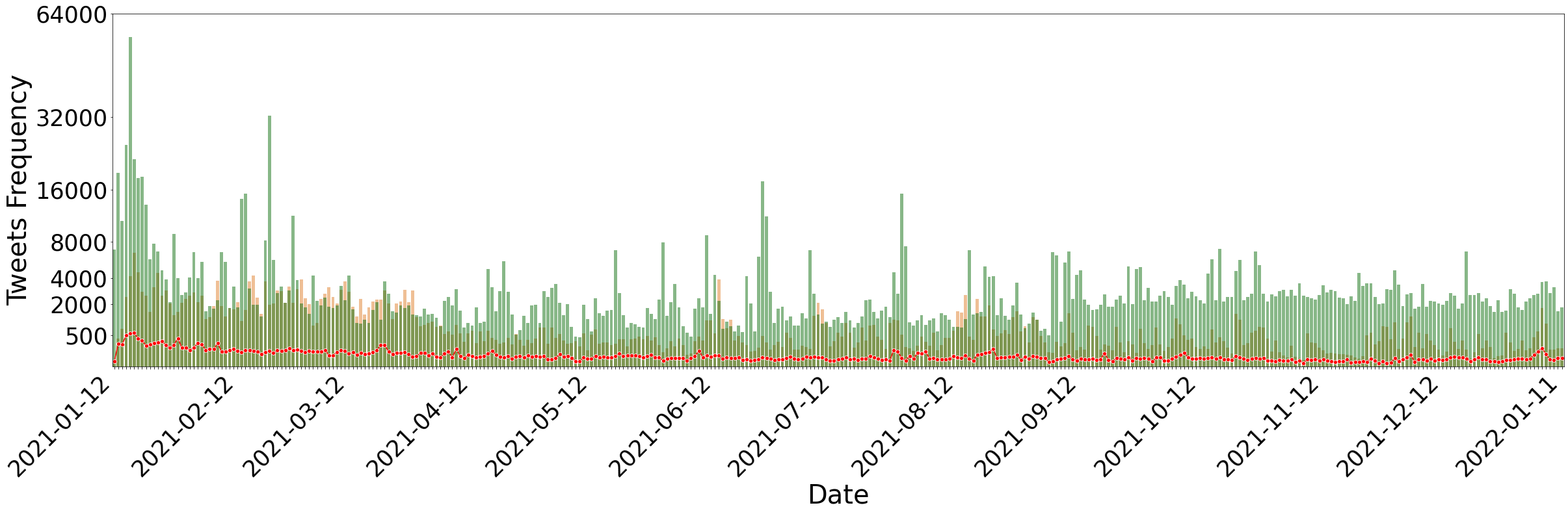}
    \vspace{-0.3cm}
    \caption{Number of tweets posted each day from 2021 Jan 12 to 2022 Jan 22. Each column represents one day. The red line represents the original tweet frequency, the orange bars represent the original tweets with retweets frequency, and the green bars represent the reply frequency. For better representation, the $y$-axis is scaled $x^{1/2}$.}
    \label{tweets_activity}
    \vspace{-0.3cm}
\end{figure*}

\section{Methods}

This section introduces our methods, which focus on assessing what topics and questions gathered the most attention, and identifying different roles within the Sedition Hunters community. 

\noindent {\textbf{Data.}}
Using the Twitter API 2.0 search endpoints, we used the hashtag \texttt{\#Seditionhunters} as query to collect all the tweets from 2021 January 12 (when the first usage appeared) to 2022 January 12 (one year later). As the Capitol Attack and discussions around it focused on the U.S., we only store tweets in English and their associated metadata. There were 321,662 unique tweets (25,324 if excluding retweets) over the year, with 65,282 users participating. 
From all collected tweets, only a fraction ($7.87\%$) are original tweets or quote tweets where new content is presented. The rest are retweets. However, the Twitter search endpoint ignores tweet replies, an important dataset consisting of many original discussions and conversations. Therefore, we further collected all replies to tweets using conversation IDs provided by the Twitter API. There are 8,317 tweets that have replies, totaling 1,173,870 replies (including replies of replies). 
We collected the data in 12 monthly intervals starting from the 12\textsuperscript{th} day of each month to the 11\textsuperscript{th} day of the following month. Tweet frequency is shown in Figure \ref{tweets_activity}.
\vspace{0.1cm}

\noindent {\textbf{Topic Model.}}
Given the large size of the dataset, manual text analysis is infeasible. Thus, we use a topic modeling algorithm to analyze what discussion took place within the \texttt{\#SeditionHunters} community. BERTopic is an algorithm that leverages BERT embeddings and dense-based clustering to overcome the limitations of other methods such as LDA. In particular, LDA, as one of the most widely used methods, assumes that documents are created from words that belong to different topics. This is usually not the case with Twitter data in which one tweet is normally about one specific topic. Also, LDA requires prior knowledge of topic numbers, which is hard to conclude in this scenario, and it makes the bag-of-words assumption, ignoring the context of words \cite{egger2021identifying, blei2003latent}. In BERTopic, the assumption of each data entry is talking about only one topic is especially convenient in that it allows us easily trace back each topic and look at the original texts of the topic. BERTopic first transforms each document into a dense vector so that documents within close spatial proximity can be grouped using the HDBSCAN \cite{mcinnes2017hdbscan} clustering algorithm. Then, topics were extracted and described by class-based TF-IDF. In addition, BERTopic uses Uniform Manifold Approximation and Projection (UMAP) \cite{mcinnes2018umap} for dimension reduction in that it often deals with high dimensional vectors. A detailed explanation is available in the Appendix.

In our study, we first filter out the retweets, since their textual content is simply repeating the original tweet. Then, we remove the usernames and hashtags from the tweet. Next, we lemmatize the words in the remaining tweets and remove the duplicate after they are processed. Finally, we pass this subset to BERTopic. For better interpretability, we fine-tune BERTopic so that the number of topics is limited to a reasonable range. Given the scale of the data, the number of topics can be reduced to around 100 while still having interpretable results. After we acquire the result, we use the top-10 most weighted topics as a guide to categorize and present other less weighted topics. We manually examine at least 100 tweets for presented topics to support our statements.
\vspace{0.1cm}

\noindent {\textbf{Network Analysis}
In addition to the text analysis with topic modeling, we employ social network analysis (SNA) to investigate the structure of the community, i.e., who participated in the \texttt{\#SeditionHunters} community, and how they participated. As mentioned, retweet activity is dominant. Therefore, we first build the retweet network where each node in the graph represents a Twitter account and directed edges between nodes represent retweets. We additionally construct a mention network where each node represents a Twitter account and directed edges represent mentioning activity. The mention network is complementary to the retweet network; a retweet activity shows where the information comes from, and the mention activity shows where the information goes. 

SNA offers numerous metrics that can show the importance of each node, and these metrics also reflect the social capital of the nodes \cite{burt1992structural}. We make use of three basic metrics: in-degree, out-degree, and modularity. In-degree refers to the number of connections received in a directed network. For the retweet network, high in-degree means the user is retweeting many other users, which implies high participation. A node with a high in-degree in the mention network is a node where many other users are directing information towards it. Out-degree is the opposite of in-degree and measures the number of connections sent in a directed network. In the retweet network, a node with a high out-degree means the respective node received many retweets. Such users usually have high visibility since the node influenced the spread of the information. Finally, the likelihood of forming clusters or groupings of closely connected nodes in a network is measured by modularity. Higher modularity means connections within a group are denser and links to other groups are less dense \cite{NewmanModularity}. When calculating modularity using Gephi \cite{bastian2009gephi}, we also compute the number of modules. A high number of modules usually implies a more disconnected network where informational flow might be negatively affected \cite{modularityInfoFlow}. 

The original dataset is processed into networks and then visualized using the Force Atlas 2 layout in Gephi \cite{Bastian_Heymann_Jacomy_2009}. We analyze the community and list the important users according to these metrics (see Tables \ref{table-degree} and \ref{table-degree-mention}). Finally, we examine the Twitter account page of important users to categorize them.

\begin{figure*}[t!]
  \centering
  \begin{minipage}[c]{0.2\textwidth}
    \caption*{Topic 0, 6.5\%}
    \vspace*{-1mm}
    \includegraphics[width=\textwidth]{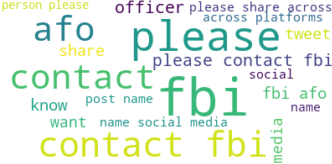}
  \end{minipage}
   \hspace{2ex}
  \begin{minipage}[c]{0.17\textwidth}
    \caption*{Topic 1, 5.8\%}
    \vspace*{-1mm}
    \includegraphics[width=\textwidth]{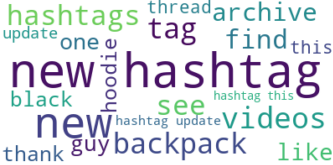}
  \end{minipage}
    \hspace{2ex}
  \begin{minipage}[c]{0.17\textwidth}
    \caption*{Topic 2, 4.6\%}
    \vspace*{-1mm}
    \includegraphics[width=\textwidth]{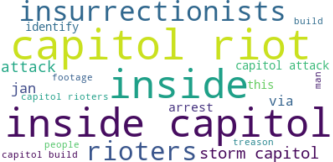}
  \end{minipage}
    \hspace{2ex}
  \begin{minipage}[c]{0.17\textwidth}
    \caption*{Topic 3, 2.8\%}
    \vspace*{-1mm}
    \includegraphics[width=\textwidth]{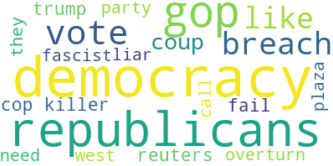}
  \end{minipage}
    \hspace{2ex}
  \begin{minipage}[c]{0.17\textwidth}
    \caption*{Topic 4, 1.5\%}
    \vspace*{-1mm}
    \includegraphics[width=\textwidth]{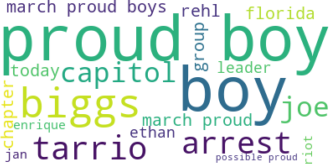}
  \end{minipage}
  \begin{minipage}[c]{0.17\textwidth}
    \caption*{Topic 5, 1.4\%}
    \vspace*{-1mm}
    \includegraphics[width=\textwidth]{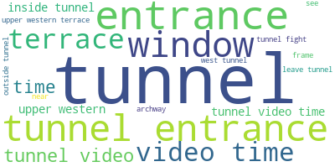}
  \end{minipage}
  \hspace{2ex}
  \begin{minipage}[c]{0.17\textwidth}
    \caption*{Topic 6, 1.3\%}
    \vspace*{-1mm}
    \includegraphics[width=\textwidth]{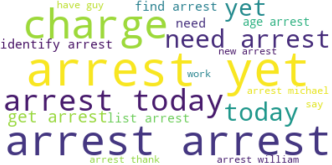}
  \end{minipage}
  \hspace{2ex}
  \begin{minipage}[c]{0.17\textwidth}
    \caption*{Topic 7, 1.0\%}
    \vspace*{-1mm}
    \includegraphics[width=\textwidth]{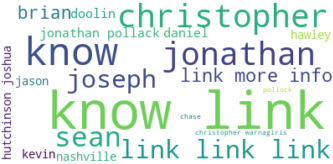}
  \end{minipage}
  \hspace{2ex}
  \begin{minipage}[c]{0.17\textwidth}
    \caption*{Topic 8, 0.9\%}
    \vspace*{-1mm}
    \includegraphics[width=\textwidth]{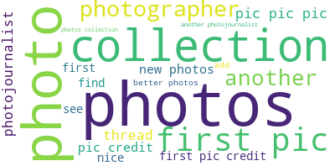}
  \end{minipage}
  \hspace{2ex}
  \begin{minipage}[c]{0.17\textwidth}
    \caption*{Topic 9, 0.9\%}
    \vspace*{-1mm}
    \includegraphics[width=\textwidth]{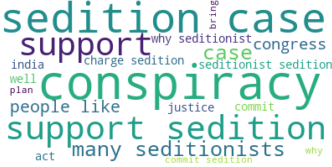}
  \end{minipage}
  \vspace{-0.2cm}
  \caption{Keywords and weights for the top 10 most weighted topics}
  \label{tweetsTopics}
\end{figure*}

\begin{figure}[t!]
    \centering
    \includegraphics[width=0.65\columnwidth]{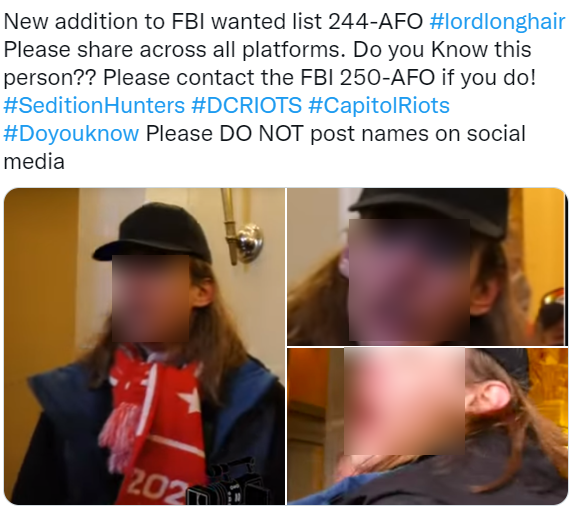}
     \vspace{-0.3cm}
    \caption{Example tweet for Topic 0 (information sharing)}
    \label{top0tweet}
    \vspace{-0.3cm}
\end{figure}

\section{Findings}

As shown in Figure~\ref{tweets_activity}, the community quickly generated much attention in the first month, with many discussions. The retweet counts remained at a high level for the first three months. They gradually decreased with small bursts in June, August, December, and the following January. The reply frequency has bursts across the year. However, unlike retweets, as the frequency decreased during the first 9 months, it increased around September 12 and remained at a relatively high level afterward.

\subsection{What are the main topics or targets of \#SH?}
Using BERTopic, we extract 119 topics from the original tweets and 57 topics from the replies. These topics reveal several main categories describing how users participated in the \texttt{\#SeditionHunters} discussion. First, a majority of the tweets are related to asking for help and promoting information sharing. Second, there are some topics related to how this community keeps track of progress. Third, there are discussions about news, events, and updates that are not necessarily related to the online sleuthing process, including many participants simply tweeting as a means of self-expression. 

\noindent \textbf{Topics from the Tweets:} Figure \ref{tweetsTopics} shows the top 10 topics out of the 119 topics from tweets in terms of their weight. Many heavily weighted topics are about information sharing and asking for information sharing. For example, Topic 0 has keywords such as "please contact fbi", "fbi afo" ("afo" refers to target labels assigned by the FBI), "please share across", and "across platform". An example tweet of Topic 0 is shown in Figure~\ref{top0tweet}. Similar tweets with different target labels are a major component of the \#SeditionHunters community. We also find other topics with similar semantic meanings but phrased differently, such as Topic 18 with keywords "track tag", "hey see", "recognize contact directly" and Topic 28 with keywords "let identify", "need identify", "recognize", and "anyone identify".

In Topic 0, the specified targets are often independent of each other. In contrast, some topics cluster tweets that have specific targets related to a single theme. Topic 4, for example, has keywords such as "proud boy", "march proud boys", "joe biggs", and "enrique tarrio". These keywords are all related to the Proud Boys, a right-wing extremist group that has taken part in multiple acts of violence and intimidation. Out of around 110 active members, more than 30 Proud Boys have been identified in connection with the Capitol Attack \cite{proudboy}. Some tweets labeled as Topic 4 are asking for more information about specific members of the Proud Boys, similar to the tweets in Topic 0. Others are news and updates related to the Proud Boys, such as the leader of the Proud Boys has been charged with conspiracy in the January 6 attack, and another leading figure of the Proud Boys who participated in the Capitol Attack and was consequently arrested later.

Topic 5 with keywords "tunnel entrance", "tunnel video", "inside tunnel", "upper western terrace", and "tunnel fight" is another single-theme topic. This topic describes the 3-hour-long fight that happened around the west terrace and inside the western tunnel entrance of the Capitol where rioters fought the police blockade and tried to gain entry to the Capitol. Comparing the tweets of Topic 0 and Topic 5, the first contains tweets asking for images of specific targets with publicized information. As we can see in Figure \ref{top0tweet}, the target is already on the FBI's wanted list with the FBI-generated suspect ID 244-AFO. However, Topic 5 has more original content such as capturing images and creating hashtags of individuals who were involved in the tunnel fight. As the whole tunnel was blocked by the police, a picture of an individual entering or leaving the tunnel can be used to accuse the person of fighting the police officers inside the tunnel. Such information is usually passed to the FBI and contributes to the formal criminal investigation process.     

Besides distributing information about potential suspects, related updates regarding legal proceedings are often shared as well. Some of the tweets in Topic 6 with keywords "arrest yet", "charge", "arrest today", together with Topic 61's keywords "update arrest", "link update arrest", "link update", show that there are constant updates on the sleuthing process.

The topic model also shows hints on how \texttt{\#SeditionHunters} keep track of the progress. Topic 1 with keywords "new hashtag", "videos", "hashtags", "tag", "find", and "archive", shows that hashtags are one way for the entire community to track specific targets. As mentioned above, some of these hashtags come from law enforcement officials' suspect IDs, while others are newly generated by Sedition Hunters. In addition to hashtags, Topic 32 with keywords such as "spreadsheet", "data", "add spreadsheet", "info sheet" reveals that the community uses shared Google Docs spreadsheets to keep track of progress. After examining the corresponding tweets and spreadsheets, we find that the usage of the spreadsheets is multi-purpose. Some spreadsheets are used to keep track of multiple targets. Each entry in such spreadsheets is about one specific target and its related information such as legal status, corresponding hashtag, appearance notes, etc. Other spreadsheets might be used to keep track of one specific video as each entry is an important frame of the video with its related information.

\begin{figure}[t!]
    \centering
    \subfloat[\centering "Perp Sheet" page of the official Sedition Hunters website]{\includegraphics[width=0.4\columnwidth]{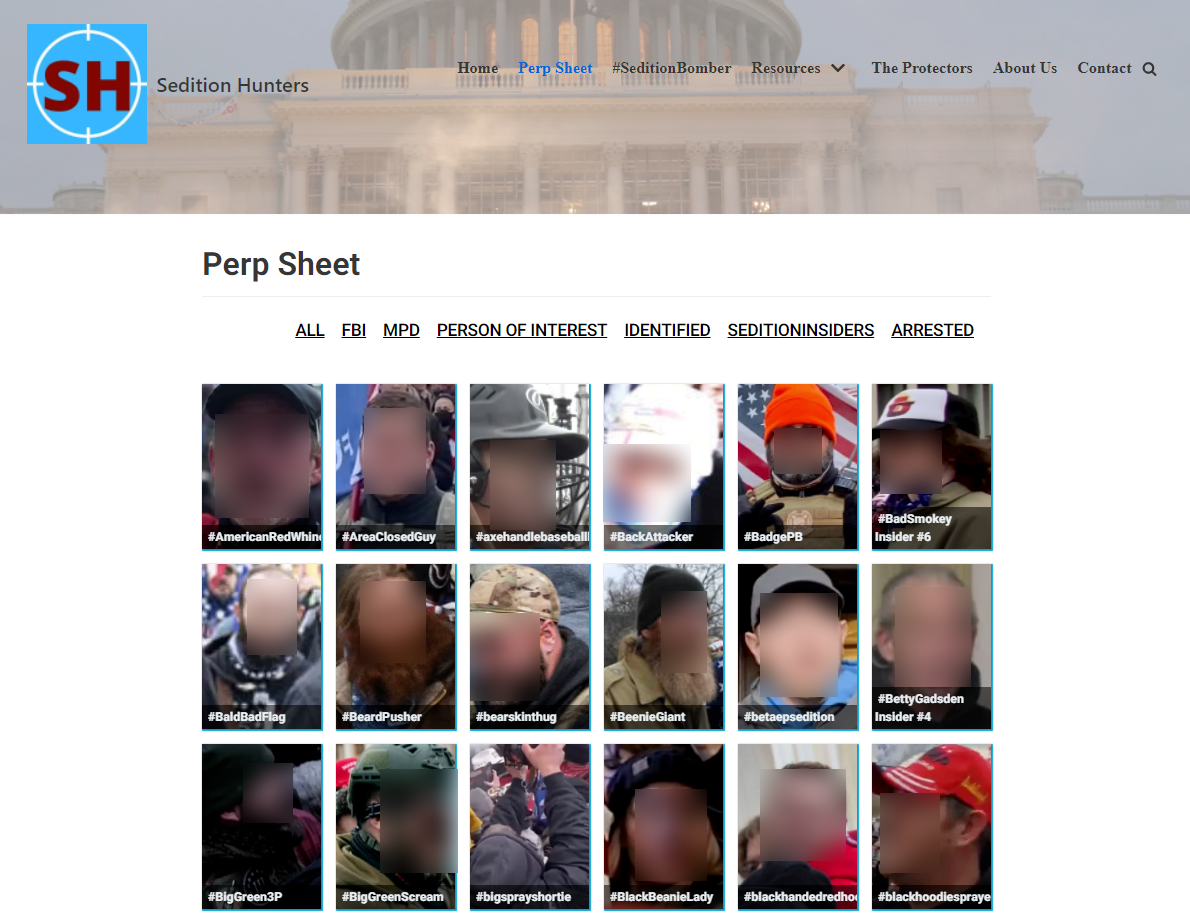}}
    \quad
    \hfill
    \subfloat[\centering Interactive map locating different evidence videos in chronological order]{
    \includegraphics[width=0.45\columnwidth]{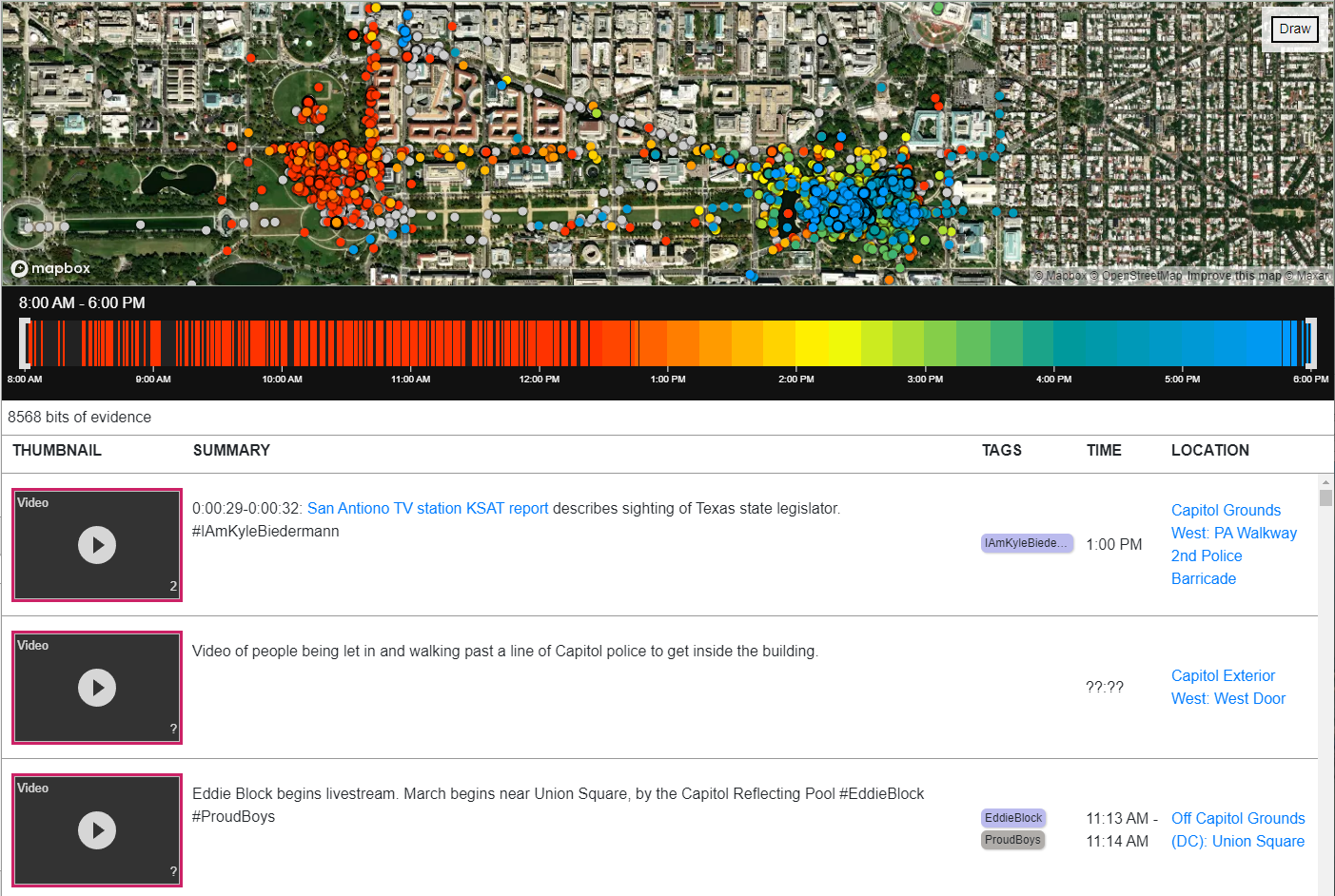}}
    \vspace{-0.3cm}
    \caption{Related websites for Sedition Hunters}
    \label{webExample}
    \vspace{-0.4cm}
\end{figure}
Another way for Sedition Hunters to track and organize their progress is by building suspect profile web pages. Topic 7 has keywords such as "know link", "know", "link more info". Back tracing the tweets and the links from Topic 7, we find several community-built websites. All of them have a listing page that tracks all targeted individuals. Although presented in different styles, these listing pages are often associated with corresponding individual profile pages which provide sleuthing progress updates. For example, the official website of \texttt{\#SeditionHunters} has a "Perp sheet" where all avatars of suspects are listed (see Figure~\ref{webExample}a). Each avatar leads to its corresponding profile page which has details such as pictures or videos of the suspect committing crimes during the Capitol Attack and corresponding legal documents. Other websites incorporate interactive tools for better usage. For example, \textit{jan6attack.com} provides detailed filters in which the user can search for a person of interest by headgear types, face covering colors, top wear brands, etc. Similar to spreadsheets, these websites are not limited to tracking suspects. The website \textit{jan6evidence.com} pinned over 8000 videos on an interactive map in chronological order (see Figure \ref{webExample}b). Moreover, \textit{capitolmap.com} has an interactive face recognition tool that takes a portrait photo and returns similar faces so that the user can easily ``compare faces'' based on the input. 

Parallel to the topics related to the online investigation, many tweets commented on discussions about trending events. For example, Topic 3 is about the discussion of "democracy", "republicans", "GOP", and "vote". The discussion may relate to the January 6 attack, but they are not directly associated with the sleuthing process. Furthermore, some topics show that some tweets are self-expressions with no clear intention of making conversation or contributing to the investigation process. For instance, Topic 13 with keywords "yeah baby", "lol", "shit", "wtf wtf"; Topic 24 with keywords "great jobs", " heroes thanks", "work good"; and a more negative Topic 60 with keywords "loser", "douchebag" and "stfu".

In addition to the three main categories mentioned above, as \cite{venkatagiri2021sedition} showed, in order to avoid ethical issues such as exposing private information,  many "DoYouKnow" composites created by the Sedition Hunters explicitly instructed community members to refrain from publicly naming suspects and instead submit tips directly to the FBI. We trace back to some of the original tweets using keywords such as "contact fbi" in Topic 0 and "recognize contact directly" and "directly" in Topic 18. We found similar textual cues, suggesting that the community is willing to learn from their own mistakes \cite{konkol_trolls_2021} and previous events \cite{lee2015realstory} so that they can minimize negative impacts caused by errors such as misidentification.

\noindent \textbf{Topics from replies:} The large number of replies led to 2165 topic clusters. For a deeper investigation, we increased the n-neighbors of the UMAP from 10 to 100. Accordingly, we increased the minimal topic size of HDBScan from 10 to 100. This setup reduced the final number of topics to 448, which are either discussions of trending events or self-expression.
For event discussion topics, most of them do not contribute to the investigation process. For example, Topic 1 is talking about "joe biden", "obama" and "worst president". Topic 3 is talking about "impeachments" and "twice impeach" of the "president". Some event discussion topics are not related to the Capitol Attack or Sedition Hunters at all. For example, Topic 5 is mainly about "taliban" and "afghanistan". Topic 8 is about the COVID with keywords such as "vaccines", "covid", and "wear mask". And Topic 9 is discussing "debt" and "tax cut". Additionally, many replies are used as a means of self-expression: Topic 4 has keywords such as "stupid", "know", "idiots", "happen", and "dumbest". 

\noindent \textbf{Outliers:} 48.3\% of tweets are classified as outliers even though the n-neighbors of the UMAP and the minimal topic size of HDBScan were set to a small number (10), which means the textual content of the community is sparse. The topic results of the replies show similar trend as 51.8\% of the replies are outliers. 
The content of the replies is less regulated compared with the tweets. This is expected as discussions can easily diverge in the reply section. Also, most replies did not contribute to the online investigation process.

The textual content does not characterize the uniqueness of this community. If we take tweets, retweets, and replies all into account, we find that information sharing and distributing tweets, especially with specified targets, are the main focus of this community. This aligns with one of the community's self-defined goals, which is creating “Do You Know” posters and composites and asking the public to share them on social media platforms \cite{seditionhunterWeb}. Among the distributed targets, there is no single target that stands out. However, some targets together lead to a converged theme such as the "tunnel fight" and the "Proud Boys". Additionally, we notice the increased awareness of protecting private information, as we see explicit efforts of the community where they request people to not use real names and contact law officials directly for providing tips.

\subsection{Who participates in \#SH and how?}
We use SNA to identify different roles in the \#SH community. We transform the original dataset into two types of networks, retweet and mention networks. In retweet networks, each node in the graph represents a Twitter account and the directed edges represent retweet activities. An edge from node A to node B represents B retweeting A. In mention networks, nodes represent Twitter accounts and edges represent mention activities where an edge from node A to node B means that A posted a tweet mentioning B.

\noindent \textbf{Retweet Network:} There are 63,859 nodes and 126,353 edges. The graph modality, 0.473, is relatively high given the range of 0 to 1. We colored the top 10 out of 107 groups according to the modality as shown in Figure~\ref{reNW1} (a). Besides the central dense area, many groups are centered around one or two high-degree nodes. For example, the nodes in the red group (23.24\% of the total number of nodes) almost all directly connect to the central node "SeditionHunters".

To better identify different roles in the community, we manually group users according to their degrees as shown in Figure~\ref{reNW1} (b). The first group is the pink nodes with a degree of 1. More than half of the nodes (62.73\%) belong to this group. They are almost exclusively in periphery positions. These nodes were pushed outward and converged as clusters because they have retweet relations with only one other node. Note that a user could retweet a second user many times without increasing the first user's degree. A degree of 1 does not necessarily imply an inactive user. Instead, it means the user did not engage comprehensively with more users in the dataset. 
The second group is yellow nodes with degree 2--10, accounting for 35.69\% of nodes. Compared to the first group, these nodes are positioned more in the center area. We also see some clusters in this group. They are formed when the nodes within are retweeting from the same nodes. In addition to the clusters, we see a more distributed pattern, implying that users may be more engaged. 
The third group is nodes in blue with degree more than 10. Only 1.58\% of total nodes have degree more than 10. However, these nodes are the backbone of the \texttt{\#SeditionHunters} community. Nodes in this group are dispersed and often are the source of the clusters. These accounts either posted original content worthy to be retweeted many times, or it is already an influential account that brings more exposure to the content, leading to more retweets.

As the degree increases, we can then focus on the differences between in-degree and out-degree. The in-degree interval of the retweet network falls between 0 and 4,487. The out-degree is between 0 and 30,429. Figure~\ref{in-out-degree} shows the high in- and out-degree users in the retweet network. The larger size and greener hue means the in- or out-degree is higher. Notice that there there is only one node that has both high in-degree and out-degree.   

\begin{figure}
\vspace{-7mm}
    \centering
    \subfloat[\centering Grouped by modality]{
    \includegraphics[width=0.38\columnwidth]{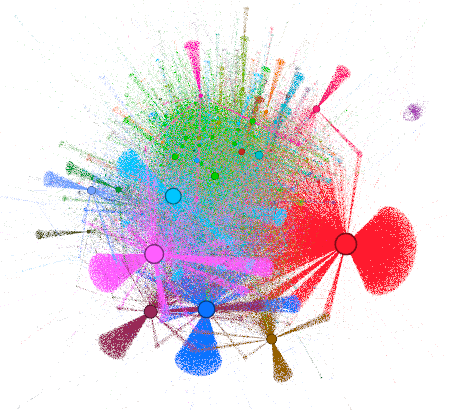}}
    \qquad
    \subfloat[\centering Grouped by degree, Pink: degree 1, Yellow: degree 2 to 10, Blue: degree more than 10]{
    \includegraphics[width=0.4\columnwidth]{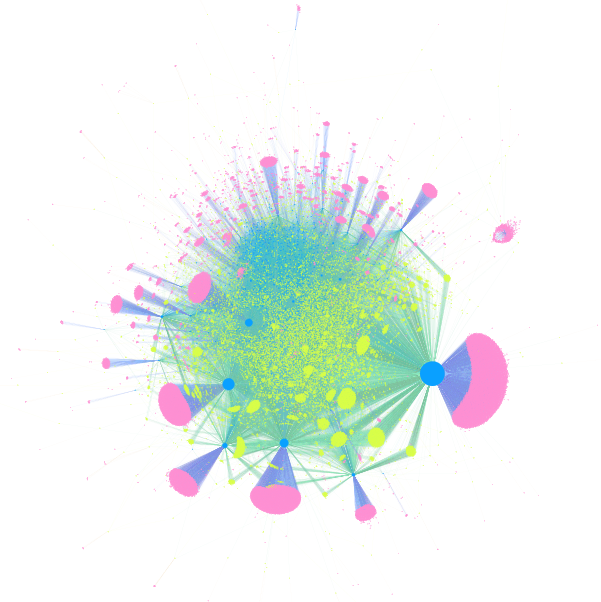}}
     \vspace{-0.3cm}
    \caption{Retweet networks grouped by modality or degree}
    \label{reNW1}
    \vspace{-0.5cm}
\end{figure}

\begin{figure}
    \centering
    \subfloat[\centering In-degree]{ \includegraphics[width=0.4\columnwidth]{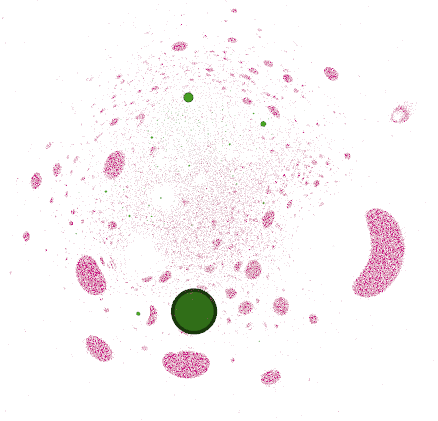}}%
    \qquad
    \subfloat[\centering Out-degree ]{\includegraphics[width=0.4\columnwidth]{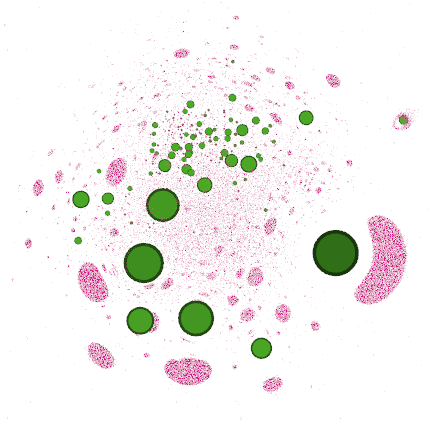}}%
     \vspace{-0.3cm}
    \caption{Retweet network node size proportional to in-degree and out-degree}%
    \label{in-out-degree}%
    \vspace{-0.3cm}
\end{figure}

To understand the different roles that users played in the large \texttt{\#SeditionHunters} community, we further analyze the top 10 users with higher in-degree and out-degree (see Table~\ref{table-degree}). As suggested above, users with higher degrees are essential to the functioning of the community. We investigate the follower account along with the degrees with the assumption that users with more followers are able to distribute information more effectively.

\begin{table*}[t]
  \centering
  {\small
  \resizebox{0.9\linewidth}{!}{
  \begin{tabular}{ p{8em} | p{5.1em} | p{5.1em} | p{7.1em} || p{8em} | p{5.1em} | p{5.1em} | p{7.1em} }
    \toprule
    \multicolumn{4}{c||}{\textbf{Top in-degree users}} &
    \multicolumn{4}{c}{\textbf{Top out-degree users}}\\
    \hline
    \textbf{Usernames} &  \textbf{Out-degree} &  \textbf{In-degree} &  \textbf{Follower counts} &   \textbf{Usernames} &  \textbf{Out-degree} &  \textbf{In-degree} &  \textbf{Follower counts}\\ 
    \hline
    
    @ryanjreilly & 10,341 & 4,487 & 167,835 & @SeditionHunters & 30,429 & 0 & 64,310 \\ 
   
    @patriottakes & 568 & 941 & 449,980 & @capitiolhunters & 14,311 & 6 & 34,763 \\
    
    @Anonymized-1 
 & 501 & 513 & 255 & @ryanjreilly & 10,341 & 4,487 & 167,835\\
    
    @DomesticTerror2 & 5,670 & 352 & 583 &  @seditiontrack & 8,886 & 181 & 70,196 \\
     
    @Anonymized-2  & 217 & 274 &  850 & @DomesticTerror2 & 5,670 & 352 & 583\\
    
    @nycjim & 126 & 258 &  221,756 & @Anonymized-3 & 3,301 & 108 & 324\\
    
    @SlickRockWeb & 1,173 & 249 & 6,951 &  @nate\_thayer & 2,252 & 30 & 6,666\\
    
    @SeditionSleuth & 88 & 209 & 495 & @Anonymized-4 & 2,143 & 10 & 8,923\\
    
    @Anonymized-5 & 154 & 197 & 31,480 & @Anonymized-6 & 1,832 & 23 & 709\\
   
    @Anonymized-7 & 486 & 188 & 545 & @Anonymized-8 & 1,679 & 22 & 305\\ 
    \bottomrule
    
    \end{tabular}}}
  \caption{Users with top in- and out-degree in retweet network (full usernames are anonymized except for high-profile accounts)}
  \vspace{-0.5cm}
  \label{table-degree}
   \vspace{-0.2cm}
\end{table*} 

For higher out-degree accounts, \texttt{@SeditionHunters} is the official account of the community. Accounts such as\texttt{@capitolhunters}, \texttt{@seditiontrack}, and \texttt{@DomesticTerror2} are created specifically for \texttt{\#SeditionHunters} shortly after January 6, 2021. \texttt{@ryanjreilly} and \texttt{@nate\_thayer} are journalists. Others are normal Twitter accounts without obvious Sedition Hunters-related identifications such as usernames, profile pictures, or profile descriptions. However, by looking at the most 100 recent tweets, we find that tweets of \texttt{@Anonymized-4} and \texttt{@Anonymized-6} are exclusively about the Capitol Attack and Sedition Hunters.

Compared with top out-degree accounts, most of the higher in-degree accounts are normal Twitter users. Besides \texttt{@ryanjreilly} and \texttt{@DomesticTerror2}, we already see in the out-degree list, many users are passionate about politics. For example, \texttt{@patriottakes} is described as an account of researchers exposing right-wing extremism, and \texttt{@Anonymized-5} self-identifies as a political activist. 
The in-degree range is considerably lower than the out-degree. The highest in-degree is 4,487, then it quickly decreases to a few hundred. In-degree indicates the number of retweets one user gathers, which might be retweeted by their own followers. Thus, these nodes can be considered information transmitters. More green nodes in the out-degree graph together with the large gap between in-degree and out-degree confirmed the idea that the main goal of the community is sharing and distributing information.

The analysis of Table~\ref{table-degree} shows that the number of followers does not reflect how important the account is. There are accounts that have few followers but high in- or out-degrees, and vice versa. 
It is also noticeable that the top out-degree nodes rarely retweet others. The highest out-degree account, @SeditionHunters, never retweets other users. This suggests that they act more as an authority or information generator. The one node with large in-degree and out-degree is \texttt{@ryanjreilly}, an NBC Justice News reporter.  

\begin{figure}[t!]
     \centering
    \includegraphics[width=0.5\columnwidth]{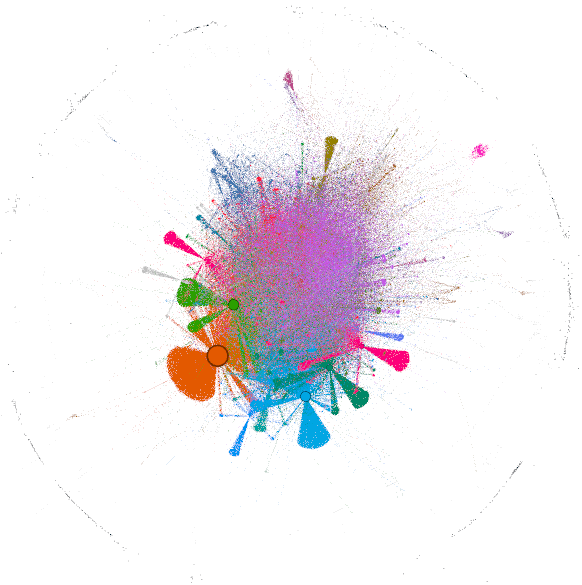}
    \vspace{-0.5cm}
    \caption{Mention Network grouped by modality}
    \label{mentionNW0}
    \vspace{-0.5cm}
\end{figure}
\noindent \textbf{Mention Network}: There are 70,317 nodes and 169,623 edges. As there are slightly more nodes and edges, the total of 734 detected communities is significantly higher than the retweet network (107 detected communities). However, the modality, 0.474, is very close to the retweet network (0.473), meaning the intensity of the connection within each group is almost identical to the retweet network, but it is more disconnected.

The mention network also has many clusters formed around the dominant accounts (see Figure~\ref{mentionNW0}). Unlike retweeted users, mentioned users do not need any related content to be mentioned. Thus, the mention network contains 0-degree accounts and they are pushed to the periphery forming the circle around the graph.  
The degree distribution also shows a similar trend in which more than half of the users (55.89\%) have a degree of 1. 39.99\% of users have degree of 2 to 10, and 3.31\% of users have degree of more than 10. As mentioned above, the mention dataset now contains 0-degree users (0.81\%). In the context of Twitter communication, the mention degree tends to be higher than the retweet degree since users can mention multiple names in one tweet, but one specific tweet can only be retweeted once. This partially explains why there are more users with a degree of more than 1. 

Table~\ref{table-degree-mention} depicts the top 10 users with in-degree and out-degree. We can see that the scale of the in-degree is remarkably larger than the out-degree. Users with high in-degree are pivotal because many other users expose and direct the information to them. Nine of the top 10 in-degree users are also on the list of top 10 out-degree users in the retweet network, which emphasizes their importance.
Many high out-degree users in the retweet network are also in the top 10 in-degree list of the mention network. In addition to the users already introduced, we see two other noteworthy accounts: \texttt{@MacfarlaneNews}, the CBS News congressional correspondent, and \texttt{@January6thCmte}, the official account of the Select Committee to investigate the January 6th attack.

\begin{table*}[t]
  \centering
  {\small
    \resizebox{0.9\linewidth}{!}{
  \begin{tabular}{ p{8em} | p{5.1em} | p{5.1em} | p{7.1em} || p{8em} | p{5.1em} | p{5.1em} | p{7.1em} }
    \toprule
    \multicolumn{4}{c||}{\textbf{Top in-degree users}} &
    \multicolumn{4}{c}{\textbf{Top out-degree users}}\\
    \hline
    \textbf{Usernames} &  \textbf{Out-degree} &  \textbf{In-degree} &  \textbf{Follower counts} &   \textbf{Usernames} &  \textbf{Out-degree} &  \textbf{In-degree} &  \textbf{Follower counts}\\ 
    \hline
    @SeditionHunters & 31,590 & 0 & 64,310 &
     @mAnonymized-1 & 1,546 & 640 & 255 \\
    
    @capitiolhunters & 15,672 & 3 & 34,763 &
    @MacFarlaneNews & 822 & 467 & 271,076 \\
    
    @ryanjreilly & 15,038 & 15 & 167,835 &
    @January6thCmte & 169 & 467 & 594,111 \\
    
    @seditiontrack & 8,966 & 207 & 70,196 &
    @Anonymized-2 & 281 & 352 & 850 \\
    
    @DomesticTerror2 & 5,939 & 119 & 583 &
    @SeditionSleuth & 103 & 276 & 495 \\
     
    @Anonymized-3 & 3,332 & 118 & 324 &
    @SlickRockWeb & 1205 & 270 & 6,951 \\
    
    @FBI & 2742 & 59 & 3,484,047 &
    @Anonymized-7 & 575 & 223 & 545 \\
    
    @Anonymized-4 & 2,388 & 10 & 8,923 &
    @seditiontrack & 8,966 & 207 & 70,196 \\
    
     @nate\_thayer & 2,261 & 39 & 6,666 &
    @Anonymized-5 & 154 & 199 & 31,480 \\
    
    @Anonymized-6 & 2,070 & 33 & 709 &
    @Anonymized-9 & 120 & 190 & 285,942 \\

    \bottomrule
    
    \end{tabular} } }
  \caption{Users with top in- out-degree in mention network (full usernames are anonymized except for high-profile accounts)}
  \vspace{-0.5cm}
  \label{table-degree-mention}
  \vspace{-0.2cm}
\end{table*}

\section{Discussion}
The Sedition Hunters community was formed on Twitter in response to the FBI seeking the public's help in identifying individuals who engaged in unlawful activities during the U.S. Capitol riot \cite{FBI_mostwanted}. As the main account \texttt{@SeditionHunters} quickly amassed many followers, the FBI maintained limited interactions with the crowd except by constantly publishing new suspects and requesting more information about the posted suspects. Similar constrained behavior was shown in the previous web-sleuthing of the Boston Marathon Bombing \cite{tapia2014crowdsourcing}. Sedition Hunters would sift through a large amount of data and send related information back to officials or simply publish them on the main Twitter account. This task-focused communication led to notable successes, as numerous arrests and legal proceedings used the investigation results from the community \cite{legal-proceeding-1,legal-proceeding-2}. Previous studies showed that the lack of interaction between investigation officials and the crowd posed a major problem \cite{tapia2014crowdsourcing,venkatagiri2021crowdsolve}. However, the Sedition Hunters community shows that the value of crowdsourced investigation can be realized even with constrained interaction or communication between law enforcement and the crowd. 
\vspace{0.1cm}

\noindent \textbf{From Topic Modeling:} To better understand this Twitter community, our first step beyond the preliminary examination was performing a content analysis of its tweets. According to the topic modeling results, the majority of the tweets were about promoting information sharing. Despite a small number of tweets sharing related news and investigation updates, most tweets were trying to expose the specified suspects more broadly by tweeting or retweeting related information, such as the "DoYouKnow" composites.

In the 2013 Boston Marathon Bombing, police asked for help identifying two potential suspects. In contrast, the Sedition Hunters community is constantly investigating dozens of suspects. The results show the importance of keeping track of the overall investigation, especially when numerous suspects were under investigation in parallel and with a such large amount of data. We see the hashtag function of Twitter was useful as it allows targets and topics to be easily documented. For example, in addition to the FBI-generated IDs (e.g., tags with prefix or suffix AFO), the community often creates hashtags of individuals based on their appearances (e.g., \texttt{\#lordlonghair} in Figure~\ref{top0tweet}). However, as the investigation progressed, we see many tweets with a long list of hashtags indicating different descriptions of the same target. Many targets had two identification hashtags, with one created by the FBI and the other created by the crowd. This may introduce inefficiency when listed randomly at the end of a tweet.

\noindent \textit{\underline{Design Recommendation 1}:} Prior research \cite{zappavigna2015searchable} shows hashtags can convey complex meanings and shape the crowd's reaction. Hashtags as documentation tool can be improved with proper curation, such as eliminating co-references and providing structural relations \cite{starbird2010tweak}. 

\noindent {\textit{\underline{Design Recommendation 2}:}} We see self-restriction efforts such as intentionally not naming suspects in public. However, there is no explicit monitoring or prevention mechanism. Mistakes with terrible consequences (i.e., \cite{lee2015realstory}) can easily happen again. Moreover, \cite{starbird2019disinformation} shows that strategic information operations are a critical concern for CSCW researchers. We see that the entire community participates around the few influential accounts, making it fertile ground for strategic information operations such as disinformation.

In addition to hashtags, our results revealed how the crowd uses various platform-independent methods to facilitate their investigation. For example, apart from spreadsheets (similar to other crowdsourcing events \cite{bow2013crowdsourcing,guntha2020lessons}), several standalone websites have been built to systematically track the investigation process. As mentioned previously, these websites show that the crowd can incorporate advanced technology such as facial recognition tools into their investigation and create convenient yet powerful interfaces to organize and visualize the entire event. Although the author of the facial recognition website limited the functionality to avoid linking the faces with named identities, considering it may cause serious misidentification, it shows that such advanced technologies are becoming more accessible to crowd workers and they are now able to harness such tools actively or passively \cite{faceRecog1,mak2021fbiUsingClues}.

\noindent \textit{\underline{Design Recommendation 3}:} We also note that almost all websites had a profile page listing the investigation progress of suspects. This seemingly universal desire for progress tracking revealed the lack of such features on online social media platforms and emphasizes the importance of information visibility.
\vspace{0.1cm}

\noindent \textbf{From Social Network Analysis:} We used SNA to identify different roles in the Sedition Hunters community. Prior work has shown three types of investigative models: top-down \cite[e.g.,][]{alcaidinho2017mobile, poelman2012if}, bottom-up \cite[e.g.,][]{erete2015engaging, israni2017snitches}, and hybrid \cite[e.g.,][]{venkatagiri2021crowdsolve}. The Sedition Hunters community is similar to top-down law enforcement-led investigations with a few key differences. First, the crowd workers are led by a few emergent leaders within the community rather than law enforcement experts. Despite the FBI hashtags appearing in some tweets, almost all tweets are in direct relation to the main \texttt{@SeditionHunters} account. The connection between crowd workers and the few dominant accounts is closer compared to law enforcement investigations. Second, the small number of influential accounts such as \texttt{@SeditionHunter} and \texttt{@seditiontracks} can be viewed as information processors that filter information from others and create composite images. This task is usually performed by law enforcement \cite{venkatagiri2021crowdsolve}. Third, unlike traditional law enforcement-led models where information flows unidirectionally from the public to law enforcement, in Sedition Hunters, information mainly flows from influential accounts. As influential accounts act as information processors through at-mentions, they also act as information generators by instigating a significant amount of retweets that increase the spread of information.

\noindent \textit{\underline{Design Recommendation 4}:} Offline crowdsourced investigations can benefit from redistributing leadership as it divides the work and facilitates better feedback from crowdworkers \cite{venkatagiri2021crowdsolve, luther2013redistributing}. We see that the success of the Sedition Hunters community also partially relies on having multiple leading accounts. These accounts dramatically increase information visibility and open more points of contact.

Sensemaking tasks consist of information gathering, creating information representations, developing insights from the representation, and creating knowledge with the insights \cite{pirolli2005sensemaking}. The efforts of Sedition Hunters can be viewed as a large-scale collective sensemaking effort where the main goal is to gather information. However, this parallelization of search tasks may lead to individuals having incomplete access to information \cite{russell2018sensemaking}. Sedition Hunters addressed this concern by incorporating the information-sharing task into their influential accounts. This includes specifying targets, sharing progress on different suspects, and other updates. We argue this is a reason why the community achieved successful results, despite being dispersed, with different groups of low-participation users gathering around several leading accounts.

\section{Conclusion}
In this work, we examine the behavior and structure of the Sedition Hunters community. We uncover three categories of discussions and find that a small number of members played important roles in the community. We also discuss the unique characteristics of the Sedition Hunters community and how they conduct investigations. Crowdsourced intelligence can be a powerful tool for solving real-world investigations, but scrutiny over their methods can highlight how to better harness the power of crowds.

\noindent \textbf{Acknowledgements}: We wish to thank Vikram Mohanty for initial data collection. This work was supported by NSF IIS-1651969.

\bibliographystyle{ACM-Reference-Format}
\bibliography{reference}


\begin{thebibliography}{57}


\ifx \showCODEN    \undefined \def \showCODEN     #1{\unskip}     \fi
\ifx \showDOI      \undefined \def \showDOI       #1{#1}\fi
\ifx \showISBNx    \undefined \def \showISBNx     #1{\unskip}     \fi
\ifx \showISBNxiii \undefined \def \showISBNxiii  #1{\unskip}     \fi
\ifx \showISSN     \undefined \def \showISSN      #1{\unskip}     \fi
\ifx \showLCCN     \undefined \def \showLCCN      #1{\unskip}     \fi
\ifx \shownote     \undefined \def \shownote      #1{#1}          \fi
\ifx \showarticletitle \undefined \def \showarticletitle #1{#1}   \fi
\ifx \showURL      \undefined \def \showURL       {\relax}        \fi
\providecommand\bibfield[2]{#2}
\providecommand\bibinfo[2]{#2}
\providecommand\natexlab[1]{#1}
\providecommand\showeprint[2][]{arXiv:#2}

\bibitem[pro(2021)]%
        {proudboy}
 \bibinfo{year}{2021}\natexlab{}.
\newblock \bibinfo{title}{Pride \& Prejudice: The Violent Evolution of the
  Proud Boys}.
\newblock
\newblock
\urldef\tempurl%
\url{https://ctc.westpoint.edu/pride-prejudice-the-violent-evolution-of-the-proud-boys/}
\showURL{%
\tempurl}


\bibitem[sed(2021)]%
        {seditionhunterWeb}
 \bibinfo{year}{2021}\natexlab{}.
\newblock \bibinfo{title}{Sedition Hunters - About Us}.
\newblock
\newblock
\urldef\tempurl%
\url{https://seditionhunters.org/aboutus/}
\showURL{%
\tempurl}


\bibitem[fac(2021)]%
        {faceRecog1}
 \bibinfo{year}{2021}\natexlab{}.
\newblock \bibinfo{title}{This Site Published Every Face From Parler's Capitol
  Riot Videos}.
\newblock
\newblock
\urldef\tempurl%
\url{https://www.wired.com/story/faces-of-the-riot-capitol-insurrection-facial-recognition/}
\showURL{%
\tempurl}


\bibitem[leg(2021a)]%
        {legal-proceeding-1}
 \bibinfo{year}{2021}\natexlab{a}.
\newblock \bibinfo{title}{United {States} of {America} v. {Chase} {Kevin}
  {Allen}}.
\newblock
\newblock
\urldef\tempurl%
\url{www.justice.gov/usao-dc/case-multi-defendant/file/1408341/download}
\showURL{%
\tempurl}


\bibitem[leg(2021b)]%
        {legal-proceeding-2}
 \bibinfo{year}{2021}\natexlab{b}.
\newblock \bibinfo{title}{United {States} of {America} v. {David} {Nicholas}
  {Dempsey}}.
\newblock
\newblock
\urldef\tempurl%
\url{www.justice.gov/usao-dc/case-multi-defendant/file/1428151/download}
\showURL{%
\tempurl}


\bibitem[FBI(2021)]%
        {FBI_mostwanted}
 \bibinfo{year}{2021}\natexlab{}.
\newblock \bibinfo{title}{U.{S}. {Capitol} {Violence} — {FBI} {Most}
  {Wanted}}.
\newblock
\newblock
\urldef\tempurl%
\url{www.fbi.gov/wanted/capitol-violence}
\showURL{%
\tempurl}


\bibitem[Alcaidinho et~al\mbox{.}(2017)]%
        {alcaidinho2017mobile}
\bibfield{author}{\bibinfo{person}{Joelle Alcaidinho}, \bibinfo{person}{Larry
  Freil}, \bibinfo{person}{Taylor Kelly}, \bibinfo{person}{Kayla Marland},
  \bibinfo{person}{Chunhui Wu}, \bibinfo{person}{Bradley Wittenbrook},
  \bibinfo{person}{Giancarlo Valentin}, {and} \bibinfo{person}{Melody
  Jackson}.} \bibinfo{year}{2017}\natexlab{}.
\newblock \showarticletitle{Mobile collaboration for human and canine police
  explosive detection teams}. In \bibinfo{booktitle}{\emph{Proceedings of the
  2017 ACM Conference on Computer Supported Cooperative Work and Social
  Computing}}. \bibinfo{pages}{925--933}.
\newblock


\bibitem[Bastian et~al\mbox{.}(2009a)]%
        {bastian2009gephi}
\bibfield{author}{\bibinfo{person}{Mathieu Bastian}, \bibinfo{person}{Sebastien
  Heymann}, {and} \bibinfo{person}{Mathieu Jacomy}.}
  \bibinfo{year}{2009}\natexlab{a}.
\newblock \showarticletitle{Gephi: an open source software for exploring and
  manipulating networks}. In \bibinfo{booktitle}{\emph{Proceedings of the
  international AAAI conference on web and social media}},
  Vol.~\bibinfo{volume}{3}. \bibinfo{pages}{361--362}.
\newblock


\bibitem[Bastian et~al\mbox{.}(2009b)]%
        {Bastian_Heymann_Jacomy_2009}
\bibfield{author}{\bibinfo{person}{Mathieu Bastian}, \bibinfo{person}{Sebastien
  Heymann}, {and} \bibinfo{person}{Mathieu Jacomy}.}
  \bibinfo{year}{2009}\natexlab{b}.
\newblock \showarticletitle{Gephi: An Open Source Software for Exploring and
  Manipulating Networks}.
\newblock \bibinfo{journal}{\emph{Proceedings of the International AAAI
  Conference on Web and Social Media}} \bibinfo{volume}{3}, \bibinfo{number}{1}
  (\bibinfo{date}{Mar.} \bibinfo{year}{2009}), \bibinfo{pages}{361--362}.
\newblock
\urldef\tempurl%
\url{https://ojs.aaai.org/index.php/ICWSM/article/view/13937}
\showURL{%
\tempurl}


\bibitem[Bergengruen and Hennigan(2021a)]%
        {Bergengruen2021theCapitol}
\bibfield{author}{\bibinfo{person}{Vera Bergengruen} {and}
  \bibinfo{person}{W.J. Hennigan}.} \bibinfo{year}{2021}\natexlab{a}.
\newblock \bibinfo{title}{The Capitol Attack Was the Most Documented Crime in
  History. Will That Ensure Justice?}
\newblock
\newblock
\urldef\tempurl%
\url{https://time.com/5953486/january-capitol-attack-investigation}
\showURL{%
\tempurl}


\bibitem[Bergengruen and Hennigan(2021b)]%
        {bergengruen_capitol_2021}
\bibfield{author}{\bibinfo{person}{Vera Bergengruen} {and}
  \bibinfo{person}{W.J. Hennigan}.} \bibinfo{year}{2021}\natexlab{b}.
\newblock \showarticletitle{The {Capitol} {Attack} {Was} the {Most}
  {Documented} {Crime} in {History}. {Will} {That} {Ensure} {Justice}?}
\newblock \bibinfo{journal}{\emph{Time}} (\bibinfo{date}{April}
  \bibinfo{year}{2021}).
\newblock
\urldef\tempurl%
\url{time.com/5953486/january-capitol-attack-investigation/}
\showURL{%
\tempurl}


\bibitem[Beyer et~al\mbox{.}(1999)]%
        {nearNeighborMeaningful}
\bibfield{author}{\bibinfo{person}{Kevin Beyer}, \bibinfo{person}{Jonathan
  Goldstein}, \bibinfo{person}{Raghu Ramakrishnan}, {and} \bibinfo{person}{Uri
  Shaft}.} \bibinfo{year}{1999}\natexlab{}.
\newblock \showarticletitle{When Is ``Nearest Neighbor'' Meaningful?}. In
  \bibinfo{booktitle}{\emph{Database Theory --- ICDT'99}},
  \bibfield{editor}{\bibinfo{person}{Catriel Beeri} {and}
  \bibinfo{person}{Peter Buneman}} (Eds.). \bibinfo{publisher}{Springer Berlin
  Heidelberg}, \bibinfo{address}{Berlin, Heidelberg},
  \bibinfo{pages}{217--235}.
\newblock
\showISBNx{978-3-540-49257-3}


\bibitem[Blei et~al\mbox{.}(2003)]%
        {blei2003latent}
\bibfield{author}{\bibinfo{person}{David~M Blei}, \bibinfo{person}{Andrew~Y
  Ng}, {and} \bibinfo{person}{Michael~I Jordan}.}
  \bibinfo{year}{2003}\natexlab{}.
\newblock \showarticletitle{Latent dirichlet allocation}.
\newblock \bibinfo{journal}{\emph{Journal of machine Learning research}}
  \bibinfo{volume}{3}, \bibinfo{number}{Jan} (\bibinfo{year}{2003}),
  \bibinfo{pages}{993--1022}.
\newblock


\bibitem[Bow et~al\mbox{.}(2013)]%
        {bow2013crowdsourcing}
\bibfield{author}{\bibinfo{person}{Hansen~C Bow}, \bibinfo{person}{Jonathan~R
  Dattilo}, \bibinfo{person}{Andrea~M Jonas}, {and}
  \bibinfo{person}{Christoph~U Lehmann}.} \bibinfo{year}{2013}\natexlab{}.
\newblock \showarticletitle{A crowdsourcing model for creating preclinical
  medical education study tools}.
\newblock \bibinfo{journal}{\emph{Academic Medicine}} \bibinfo{volume}{88},
  \bibinfo{number}{6} (\bibinfo{year}{2013}), \bibinfo{pages}{766--770}.
\newblock


\bibitem[Burt(1992)]%
        {burt1992structural}
\bibfield{author}{\bibinfo{person}{R.S. Burt}.}
  \bibinfo{year}{1992}\natexlab{}.
\newblock \bibinfo{booktitle}{\emph{Structural Holes: The Social Structure of
  Competition}}.
\newblock \bibinfo{publisher}{Harvard University Press}.
\newblock
\showISBNx{9780674843721}
\showLCCN{lc91043396}
\urldef\tempurl%
\url{https://books.google.com/books?id=\_gjtAAAAMAAJ}
\showURL{%
\tempurl}


\bibitem[Chappell(2021)]%
        {chappell2021architect}
\bibfield{author}{\bibinfo{person}{Bill Chappell}.}
  \bibinfo{year}{2021}\natexlab{}.
\newblock \bibinfo{title}{Architect Of The Capitol Outlines \$30 Million In
  Damages From Pro-Trump Riot}.
\newblock
\newblock
\urldef\tempurl%
\url{https://www.npr.org/sections/insurrection-at-the-capitol/2021/02/24/970977612/architect-of-the-capitol-outlines-30-million-in-damages-from-pro-trump-riot}
\showURL{%
\tempurl}


\bibitem[Cochrane and Broadwater(2021)]%
        {cochrane_capitol_2021}
\bibfield{author}{\bibinfo{person}{Emily Cochrane} {and} \bibinfo{person}{Luke
  Broadwater}.} \bibinfo{year}{2021}\natexlab{}.
\newblock \showarticletitle{Capitol {Riot} {Costs} {Will} {Exceed} \$30
  {Million}, {Official} {Tells} {Congress} - {The} {New} {York} {Times}}.
\newblock \bibinfo{journal}{\emph{New York Times}} (\bibinfo{date}{Feb.}
  \bibinfo{year}{2021}).
\newblock
\urldef\tempurl%
\url{www.nytimes.com/2021/02/24/us/politics/capitol-riot-damage.html}
\showURL{%
\tempurl}


\bibitem[Daniel(2014)]%
        {daniel2014police}
\bibfield{author}{\bibinfo{person}{Trottier Daniel}.}
  \bibinfo{year}{2014}\natexlab{}.
\newblock \showarticletitle{Police and user-led investigations on social
  media}.
\newblock \bibinfo{journal}{\emph{Journal of law, information and science}}
  \bibinfo{volume}{23}, \bibinfo{number}{1} (\bibinfo{year}{2014}),
  \bibinfo{pages}{75--96}.
\newblock


\bibitem[Doig(2022)]%
        {capitolriotsize}
\bibfield{author}{\bibinfo{person}{Steve Doig}.}
  \bibinfo{year}{2022}\natexlab{}.
\newblock \bibinfo{title}{It is difficult, if not impossible, to estimate the
  size of the crowd that stormed Capitol Hill}.
\newblock
\newblock
\urldef\tempurl%
\url{https://theconversation.com/it-is-difficult-if-not-impossible-to-estimate-the-size-of-the-crowd-that-stormed-capitol-hill-152889}
\showURL{%
\tempurl}


\bibitem[Egger and Yu(2021)]%
        {egger2021identifying}
\bibfield{author}{\bibinfo{person}{Roman Egger} {and} \bibinfo{person}{Joanne
  Yu}.} \bibinfo{year}{2021}\natexlab{}.
\newblock \showarticletitle{Identifying hidden semantic structures in Instagram
  data: a topic modelling comparison}.
\newblock \bibinfo{journal}{\emph{Tourism Review}} (\bibinfo{year}{2021}).
\newblock


\bibitem[Egger and Yu(2022)]%
        {topicmodelComparison}
\bibfield{author}{\bibinfo{person}{Roman Egger} {and} \bibinfo{person}{Joanne
  Yu}.} \bibinfo{year}{2022}\natexlab{}.
\newblock \showarticletitle{A Topic Modeling Comparison Between LDA, NMF,
  Top2Vec, and BERTopic to Demystify Twitter Posts}.
\newblock \bibinfo{journal}{\emph{Frontiers in Sociology}}  \bibinfo{volume}{7}
  (\bibinfo{date}{05} \bibinfo{year}{2022}).
\newblock
\urldef\tempurl%
\url{https://doi.org/10.3389/fsoc.2022.886498}
\showDOI{\tempurl}


\bibitem[Erete(2015)]%
        {erete2015engaging}
\bibfield{author}{\bibinfo{person}{Sheena~L Erete}.}
  \bibinfo{year}{2015}\natexlab{}.
\newblock \showarticletitle{Engaging around neighborhood issues: How online
  communication affects offline behavior}. In
  \bibinfo{booktitle}{\emph{Proceedings of the 18th ACM Conference on Computer
  Supported Cooperative Work \& Social Computing}}.
  \bibinfo{pages}{1590--1601}.
\newblock


\bibitem[Ericson and Haggerty(1997)]%
        {ericson1997policing}
\bibfield{author}{\bibinfo{person}{Richard~V Ericson} {and}
  \bibinfo{person}{Kevin~D Haggerty}.} \bibinfo{year}{1997}\natexlab{}.
\newblock \showarticletitle{Policing the risk society}.
\newblock  (\bibinfo{year}{1997}).
\newblock


\bibitem[Ester et~al\mbox{.}(1996)]%
        {ester1996density}
\bibfield{author}{\bibinfo{person}{Martin Ester}, \bibinfo{person}{Hans-Peter
  Kriegel}, \bibinfo{person}{J{\"o}rg Sander}, \bibinfo{person}{Xiaowei Xu},
  {et~al\mbox{.}}} \bibinfo{year}{1996}\natexlab{}.
\newblock \showarticletitle{A density-based algorithm for discovering clusters
  in large spatial databases with noise.}. In \bibinfo{booktitle}{\emph{kdd}},
  Vol.~\bibinfo{volume}{96}. \bibinfo{pages}{226--231}.
\newblock


\bibitem[Guntha et~al\mbox{.}(2020)]%
        {guntha2020lessons}
\bibfield{author}{\bibinfo{person}{Ramesh Guntha},
  \bibinfo{person}{Sethuraman~N Rao}, {and} \bibinfo{person}{Avinash Shivdas}.}
  \bibinfo{year}{2020}\natexlab{}.
\newblock \showarticletitle{Lessons learned from deploying crowdsourced
  technology for disaster relief during Kerala floods}.
\newblock \bibinfo{journal}{\emph{Procedia Computer Science}}
  \bibinfo{volume}{171} (\bibinfo{year}{2020}), \bibinfo{pages}{2410--2419}.
\newblock


\bibitem[Israni et~al\mbox{.}(2017)]%
        {israni2017snitches}
\bibfield{author}{\bibinfo{person}{Aarti Israni}, \bibinfo{person}{Sheena
  Erete}, {and} \bibinfo{person}{Che~L Smith}.}
  \bibinfo{year}{2017}\natexlab{}.
\newblock \showarticletitle{Snitches, trolls, and social norms: Unpacking
  perceptions of social media use for crime prevention}. In
  \bibinfo{booktitle}{\emph{Proceedings of the 2017 ACM Conference on Computer
  Supported Cooperative Work and Social Computing}}.
  \bibinfo{pages}{1193--1209}.
\newblock


\bibitem[Kelling and Moore(1989)]%
        {kelling1989evolving}
\bibfield{author}{\bibinfo{person}{George~L Kelling} {and}
  \bibinfo{person}{Mark~Harrison Moore}.} \bibinfo{year}{1989}\natexlab{}.
\newblock \bibinfo{booktitle}{\emph{The evolving strategy of policing}}.
\newblock Number~4. \bibinfo{publisher}{US Department of Justice, Office of
  Justice Programs, National Institute of Justice}.
\newblock


\bibitem[Konkol(2021)]%
        {konkol_trolls_2021}
\bibfield{author}{\bibinfo{person}{Mark Konkol}.}
  \bibinfo{year}{2021}\natexlab{}.
\newblock \showarticletitle{Trolls {Wrongly} {Accused} {Retired} {Firefighter}
  {Of} {Capitol} {Riot} {Murder}}.
\newblock \bibinfo{journal}{\emph{Chicago, IL Patch}} (\bibinfo{date}{Jan.}
  \bibinfo{year}{2021}).
\newblock
\urldef\tempurl%
\url{patch.com/illinois/chicago/trolls-wrongly-accused-retired-firefighter-capitol-riot-murder}
\showURL{%
\tempurl}


\bibitem[Lee(2015)]%
        {lee2015realstory}
\bibfield{author}{\bibinfo{person}{Traci~G. Lee}.}
  \bibinfo{year}{2015}\natexlab{}.
\newblock \bibinfo{title}{The Real Story of Sunil Tripathi, the Boston Bomber
  Who Wasn't}.
\newblock
\newblock
\urldef\tempurl%
\url{https://www.nbcnews.com/news/asian-america/wrongly-accused-boston-bombing-sunil-tripathys-story-now-being-told-n373141}
\showURL{%
\tempurl}


\bibitem[Luther et~al\mbox{.}(2013)]%
        {luther2013redistributing}
\bibfield{author}{\bibinfo{person}{Kurt Luther}, \bibinfo{person}{Casey
  Fiesler}, {and} \bibinfo{person}{Amy Bruckman}.}
  \bibinfo{year}{2013}\natexlab{}.
\newblock \showarticletitle{Redistributing leadership in online creative
  collaboration}. In \bibinfo{booktitle}{\emph{Proceedings of the 2013
  conference on Computer supported cooperative work}}.
  \bibinfo{pages}{1007--1022}.
\newblock


\bibitem[Mak(2021)]%
        {mak2021fbiUsingClues}
\bibfield{author}{\bibinfo{person}{Tim Mak}.} \bibinfo{year}{2021}\natexlab{}.
\newblock \bibinfo{title}{The FBI Keeps Using Clues From Volunteer Sleuths To
  Find The Jan. 6 Capitol Rioters}.
\newblock
\newblock
\urldef\tempurl%
\url{https://www.npr.org/2021/08/18/1028527768/the-fbi-keeps-using-clues-from-volunteer-sleuths-to-find-the-jan-6-capitol-riote}
\showURL{%
\tempurl}


\bibitem[Marwick(2012)]%
        {marwick2012public}
\bibfield{author}{\bibinfo{person}{Alice Marwick}.}
  \bibinfo{year}{2012}\natexlab{}.
\newblock \showarticletitle{The public domain: Surveillance in everyday life}.
\newblock \bibinfo{journal}{\emph{Surveillance \& Society}}
  \bibinfo{volume}{9}, \bibinfo{number}{4} (\bibinfo{year}{2012}),
  \bibinfo{pages}{378--393}.
\newblock


\bibitem[Marx(2013)]%
        {marx2013public}
\bibfield{author}{\bibinfo{person}{Gary~T Marx}.}
  \bibinfo{year}{2013}\natexlab{}.
\newblock \showarticletitle{The public as partner? Technology can make us
  auxiliaries as well as vigilantes}.
\newblock \bibinfo{journal}{\emph{IEEE Security \& Privacy}}
  \bibinfo{volume}{11}, \bibinfo{number}{5} (\bibinfo{year}{2013}),
  \bibinfo{pages}{56--61}.
\newblock


\bibitem[McInnes et~al\mbox{.}(2017)]%
        {mcinnes2017hdbscan}
\bibfield{author}{\bibinfo{person}{Leland McInnes}, \bibinfo{person}{John
  Healy}, {and} \bibinfo{person}{Steve Astels}.}
  \bibinfo{year}{2017}\natexlab{}.
\newblock \showarticletitle{hdbscan: Hierarchical density based clustering.}
\newblock \bibinfo{journal}{\emph{J. Open Source Softw.}} \bibinfo{volume}{2},
  \bibinfo{number}{11} (\bibinfo{year}{2017}), \bibinfo{pages}{205}.
\newblock


\bibitem[McInnes et~al\mbox{.}(2018)]%
        {mcinnes2018umap}
\bibfield{author}{\bibinfo{person}{Leland McInnes}, \bibinfo{person}{John
  Healy}, {and} \bibinfo{person}{James Melville}.}
  \bibinfo{year}{2018}\natexlab{}.
\newblock \showarticletitle{Umap: Uniform manifold approximation and projection
  for dimension reduction}.
\newblock \bibinfo{journal}{\emph{arXiv preprint arXiv:1802.03426}}
  (\bibinfo{year}{2018}).
\newblock


\bibitem[Newman(2006)]%
        {NewmanModularity}
\bibfield{author}{\bibinfo{person}{M.~E.~J. Newman}.}
  \bibinfo{year}{2006}\natexlab{}.
\newblock \showarticletitle{Modularity and community structure in networks}.
\newblock \bibinfo{journal}{\emph{Proceedings of the National Academy of
  Sciences}} \bibinfo{volume}{103}, \bibinfo{number}{23}
  (\bibinfo{year}{2006}), \bibinfo{pages}{8577--8582}.
\newblock
\urldef\tempurl%
\url{https://doi.org/10.1073/pnas.0601602103}
\showDOI{\tempurl}
\showeprint{https://www.pnas.org/doi/pdf/10.1073/pnas.0601602103}


\bibitem[Nhan et~al\mbox{.}(2017)]%
        {nhan2017digilantism}
\bibfield{author}{\bibinfo{person}{Johnny Nhan}, \bibinfo{person}{Laura Huey},
  {and} \bibinfo{person}{Ryan Broll}.} \bibinfo{year}{2017}\natexlab{}.
\newblock \showarticletitle{Digilantism: An analysis of crowdsourcing and the
  Boston marathon bombings}.
\newblock \bibinfo{journal}{\emph{The British journal of criminology}}
  \bibinfo{volume}{57}, \bibinfo{number}{2} (\bibinfo{year}{2017}),
  \bibinfo{pages}{341--361}.
\newblock


\bibitem[Palen and Vieweg(2008)]%
        {palen2008emergence}
\bibfield{author}{\bibinfo{person}{Leysia Palen} {and} \bibinfo{person}{Sarah
  Vieweg}.} \bibinfo{year}{2008}\natexlab{}.
\newblock \showarticletitle{The emergence of online widescale interaction in
  unexpected events: assistance, alliance \& retreat}. In
  \bibinfo{booktitle}{\emph{Proceedings of the 2008 ACM conference on Computer
  supported cooperative work}}. \bibinfo{pages}{117--126}.
\newblock


\bibitem[Pirolli and Card(2005)]%
        {pirolli2005sensemaking}
\bibfield{author}{\bibinfo{person}{Peter Pirolli} {and} \bibinfo{person}{Stuart
  Card}.} \bibinfo{year}{2005}\natexlab{}.
\newblock \showarticletitle{The sensemaking process and leverage points for
  analyst technology as identified through cognitive task analysis}. In
  \bibinfo{booktitle}{\emph{Proceedings of international conference on
  intelligence analysis}}, Vol.~\bibinfo{volume}{5}. McLean, VA, USA,
  \bibinfo{pages}{2--4}.
\newblock


\bibitem[Poelman et~al\mbox{.}(2012)]%
        {poelman2012if}
\bibfield{author}{\bibinfo{person}{Ronald Poelman}, \bibinfo{person}{Oytun
  Akman}, \bibinfo{person}{Stephan Lukosch}, {and} \bibinfo{person}{Pieter
  Jonker}.} \bibinfo{year}{2012}\natexlab{}.
\newblock \showarticletitle{As if being there: mediated reality for crime scene
  investigation}. In \bibinfo{booktitle}{\emph{Proceedings of the ACM 2012
  conference on computer supported cooperative work}}.
  \bibinfo{pages}{1267--1276}.
\newblock


\bibitem[Popli and Zorthian(2022)]%
        {whatHappenedJan6}
\bibfield{author}{\bibinfo{person}{Nik Popli} {and} \bibinfo{person}{Julia
  Zorthian}.} \bibinfo{year}{2022}\natexlab{}.
\newblock \bibinfo{title}{What Happened to the Jan. 6 Insurrectionists Arrested
  Since the Capitol Riot}.
\newblock
\newblock
\urldef\tempurl%
\url{https://time.com/6133336/jan-6-capitol-riot-arrests-sentences/}
\showURL{%
\tempurl}


\bibitem[Reilly(2021)]%
        {reilly2021sedition}
\bibfield{author}{\bibinfo{person}{Ryan~J. Reilly}.}
  \bibinfo{year}{2021}\natexlab{}.
\newblock \bibinfo{title}{‘Sedition Hunters’: Meet The Online Sleuths
  Aiding The FBI’s Capitol Manhunt}.
\newblock
\newblock
\urldef\tempurl%
\url{https://www.huffpost.com/entry/sedition-hunters-fbi-capitol-attack-manhunt-online-sleuths_n_60479dd7c5b653040034f749}
\showURL{%
\tempurl}


\bibitem[Reimers and Gurevych(2019)]%
        {reimers2019sentence}
\bibfield{author}{\bibinfo{person}{Nils Reimers} {and} \bibinfo{person}{Iryna
  Gurevych}.} \bibinfo{year}{2019}\natexlab{}.
\newblock \showarticletitle{Sentence-bert: Sentence embeddings using siamese
  bert-networks}.
\newblock \bibinfo{journal}{\emph{arXiv preprint arXiv:1908.10084}}
  (\bibinfo{year}{2019}).
\newblock


\bibitem[Rizza et~al\mbox{.}(2014)]%
        {rizza2014yourself}
\bibfield{author}{\bibinfo{person}{Caroline Rizza},
  \bibinfo{person}{{\^A}ngela~Guimar{\~a}es Pereira}, {and}
  \bibinfo{person}{Paula Curvelo}.} \bibinfo{year}{2014}\natexlab{}.
\newblock \showarticletitle{“Do-it-yourself justice”: considerations of
  social media use in a crisis situation: the case of the 2011 Vancouver
  riots}.
\newblock \bibinfo{journal}{\emph{International Journal of Information Systems
  for Crisis Response and Management (IJISCRAM)}} \bibinfo{volume}{6},
  \bibinfo{number}{4} (\bibinfo{year}{2014}), \bibinfo{pages}{42--59}.
\newblock


\bibitem[Russell et~al\mbox{.}(2018)]%
        {russell2018sensemaking}
\bibfield{author}{\bibinfo{person}{Daniel~M Russell}, \bibinfo{person}{Gregorio
  Convertino}, \bibinfo{person}{Aniket Kittur}, \bibinfo{person}{Peter
  Pirolli}, {and} \bibinfo{person}{Elizabeth~Anne Watkins}.}
  \bibinfo{year}{2018}\natexlab{}.
\newblock \showarticletitle{Sensemaking in a senseless world: 2018 workshop
  abstract}. In \bibinfo{booktitle}{\emph{Extended Abstracts of the 2018 CHI
  Conference on Human Factors in Computing Systems}}. \bibinfo{pages}{1--7}.
\newblock


\bibitem[Schneider and Trottier(2013)]%
        {schneider2013social}
\bibfield{author}{\bibinfo{person}{Christopher~J Schneider} {and}
  \bibinfo{person}{Daniel Trottier}.} \bibinfo{year}{2013}\natexlab{}.
\newblock \showarticletitle{Social media and the 2011 Vancouver riot}.
\newblock In \bibinfo{booktitle}{\emph{40th anniversary of studies in symbolic
  interaction}}. \bibinfo{publisher}{Emerald Group Publishing Limited}.
\newblock


\bibitem[Smallridge et~al\mbox{.}(2016)]%
        {smallridge2016understanding}
\bibfield{author}{\bibinfo{person}{Joshua Smallridge}, \bibinfo{person}{Philip
  Wagner}, {and} \bibinfo{person}{Justin~N Crowl}.}
  \bibinfo{year}{2016}\natexlab{}.
\newblock \showarticletitle{Understanding cyber-vigilantism: A conceptual
  framework.}
\newblock \bibinfo{journal}{\emph{Journal of Theoretical \& Philosophical
  Criminology}} \bibinfo{volume}{8}, \bibinfo{number}{1}
  (\bibinfo{year}{2016}).
\newblock


\bibitem[Starbird et~al\mbox{.}(2019)]%
        {starbird2019disinformation}
\bibfield{author}{\bibinfo{person}{Kate Starbird}, \bibinfo{person}{Ahmer
  Arif}, {and} \bibinfo{person}{Tom Wilson}.} \bibinfo{year}{2019}\natexlab{}.
\newblock \showarticletitle{Disinformation as collaborative work: Surfacing the
  participatory nature of strategic information operations}.
\newblock \bibinfo{journal}{\emph{Proceedings of the ACM on Human-Computer
  Interaction}} \bibinfo{volume}{3}, \bibinfo{number}{CSCW}
  (\bibinfo{year}{2019}), \bibinfo{pages}{1--26}.
\newblock


\bibitem[Starbird and Stamberger(2010)]%
        {starbird2010tweak}
\bibfield{author}{\bibinfo{person}{Kate Starbird} {and}
  \bibinfo{person}{Jeannie~A Stamberger}.} \bibinfo{year}{2010}\natexlab{}.
\newblock \showarticletitle{Tweak the tweet: Leveraging microblogging
  proliferation with a prescriptive syntax to support citizen reporting.}. In
  \bibinfo{booktitle}{\emph{ISCRAM}}.
\newblock


\bibitem[Tapia and LaLone(2014)]%
        {tapia2014crowdsourcing}
\bibfield{author}{\bibinfo{person}{Andrea~H Tapia} {and}
  \bibinfo{person}{Nicolas~J LaLone}.} \bibinfo{year}{2014}\natexlab{}.
\newblock \showarticletitle{Crowdsourcing investigations: Crowd participation
  in identifying the bomb and bomber from the Boston marathon bombing}.
\newblock \bibinfo{journal}{\emph{International Journal of Information Systems
  for Crisis Response and Management (IJISCRAM)}} \bibinfo{volume}{6},
  \bibinfo{number}{4} (\bibinfo{year}{2014}), \bibinfo{pages}{60--75}.
\newblock


\bibitem[Van~der Maaten and Hinton(2008)]%
        {van2008visualizing}
\bibfield{author}{\bibinfo{person}{Laurens Van~der Maaten} {and}
  \bibinfo{person}{Geoffrey Hinton}.} \bibinfo{year}{2008}\natexlab{}.
\newblock \showarticletitle{Visualizing data using t-SNE.}
\newblock \bibinfo{journal}{\emph{Journal of machine learning research}}
  \bibinfo{volume}{9}, \bibinfo{number}{11} (\bibinfo{year}{2008}).
\newblock


\bibitem[Venkatagiri et~al\mbox{.}(2021a)]%
        {venkatagiri2021crowdsolve}
\bibfield{author}{\bibinfo{person}{Sukrit Venkatagiri}, \bibinfo{person}{Aakash
  Gautam}, {and} \bibinfo{person}{Kurt Luther}.}
  \bibinfo{year}{2021}\natexlab{a}.
\newblock \showarticletitle{Crowdsolve: Managing tensions in an expert-led
  crowdsourced investigation}.
\newblock \bibinfo{journal}{\emph{Proceedings of the ACM on Human-Computer
  Interaction}} \bibinfo{volume}{5}, \bibinfo{number}{CSCW1}
  (\bibinfo{year}{2021}), \bibinfo{pages}{1--30}.
\newblock


\bibitem[Venkatagiri et~al\mbox{.}(2021b)]%
        {venkatagiri2021sedition}
\bibfield{author}{\bibinfo{person}{Sukrit Venkatagiri},
  \bibinfo{person}{Tianjiao Yu}, \bibinfo{person}{Vikram Mohanty}, {and}
  \bibinfo{person}{Kurt Luther}.} \bibinfo{year}{2021}\natexlab{b}.
\newblock \showarticletitle{Sedition Hunters: Countering Extremism through
  Collective Action}. In \bibinfo{booktitle}{\emph{CSCW 2021 Workshop on
  Addressing Challenges and Opportunities in Online Extremism Research: An
  Interdisciplinary Perspective}}.
\newblock


\bibitem[Walsh and O'Connor(2019)]%
        {walsh2019social}
\bibfield{author}{\bibinfo{person}{James~P Walsh} {and}
  \bibinfo{person}{Christopher O'Connor}.} \bibinfo{year}{2019}\natexlab{}.
\newblock \showarticletitle{Social media and policing: A review of recent
  research}.
\newblock \bibinfo{journal}{\emph{Sociology compass}} \bibinfo{volume}{13},
  \bibinfo{number}{1} (\bibinfo{year}{2019}), \bibinfo{pages}{e12648}.
\newblock


\bibitem[Wojcieszak and Mutz(2009)]%
        {modularityInfoFlow}
\bibfield{author}{\bibinfo{person}{Magdalena Wojcieszak} {and}
  \bibinfo{person}{Diana Mutz}.} \bibinfo{year}{2009}\natexlab{}.
\newblock \showarticletitle{Online Groups and Political Discourse: Do Online
  Discussion Spaces Facilitate Exposure to Political Disagreement?}
\newblock \bibinfo{journal}{\emph{Journal of Communication}}
  \bibinfo{volume}{59} (\bibinfo{date}{03} \bibinfo{year}{2009}),
  \bibinfo{pages}{40 -- 56}.
\newblock
\urldef\tempurl%
\url{https://doi.org/10.1111/j.1460-2466.2008.01403.x}
\showDOI{\tempurl}


\bibitem[Yardley et~al\mbox{.}(2018)]%
        {yardley2018s}
\bibfield{author}{\bibinfo{person}{Elizabeth Yardley}, \bibinfo{person}{Adam
  George~Thomas Lynes}, \bibinfo{person}{David Wilson}, {and}
  \bibinfo{person}{Emma Kelly}.} \bibinfo{year}{2018}\natexlab{}.
\newblock \showarticletitle{What’s the deal with ‘websleuthing’? News
  media representations of amateur detectives in networked spaces}.
\newblock \bibinfo{journal}{\emph{Crime, Media, Culture}} \bibinfo{volume}{14},
  \bibinfo{number}{1} (\bibinfo{year}{2018}), \bibinfo{pages}{81--109}.
\newblock


\bibitem[Zappavigna(2015)]%
        {zappavigna2015searchable}
\bibfield{author}{\bibinfo{person}{Michele Zappavigna}.}
  \bibinfo{year}{2015}\natexlab{}.
\newblock \showarticletitle{Searchable talk: The linguistic functions of
  hashtags}.
\newblock \bibinfo{journal}{\emph{Social Semiotics}} \bibinfo{volume}{25},
  \bibinfo{number}{3} (\bibinfo{year}{2015}), \bibinfo{pages}{274--291}.
\newblock


\end{thebibliography}

\appendix
\section{Appendix}

\textbf{BERTopic:} There are four main components to the BERTopic. First, using Sentence-BERT \cite{reimers2019sentence}, each document is transformed into a dense vector representation. These vectors make it possible to locate semantically similar words, sentences, or documents within close spatial proximity \cite{topicmodelComparison}. The default all-MiniLM-L6-v2 embedding from the Sentence-Transformers package is an all-around model tuned for many use cases. However, other language models can be used for embedding. For example, BERTweet is a good alternative, trained exclusively on tweets. We chose to use the default all-MiniLM-L6-v2 embedding since it produced a lower topic coherence score (-6.06 vs. -6.76).

As data increases in dimensionality, the algorithm requires higher computational power, and more importantly, spatial locality becomes ill-defined \cite{nearNeighborMeaningful}. Thus, the next component reduces the dimensions of the embeddings. BERTopic uses Uniform Manifold Approximation and Projection (UMAP) \cite{mcinnes2018umap} for dimension reduction. Compared with other popular methods such as t-SNE\cite{van2008visualizing}, UMAP is able to leverage both local structure and global structure and be explicitly controlled by user-defined parameters. It is also faster as it creates estimations of the high dimensional graph instead of measuring every point. The core idea of UMAP is to construct a high-dimensional graph representation of the data and then optimize a low-dimensional graph that is as structurally similar as possible. In order to build the high-dimensional graph, UMAP uses simplices, which enable us to build a high-dimensional graph robustly. Starting from extending the radius of each point, UMAP connects points where the radii overlap. Choosing the radius is critical as a small radius can lead to trivial isolated clusters, while a large radius will connect everything together. UMAP overcomes this by choosing the radius locally. However, as the distance within the radius increases, the probability of connecting the original points with other points decreases \cite{mcinnes2018umap}. 

The third component is hierarchical density-based spatial clustering of applications with noise (HDBSCAN \cite{mcinnes2017hdbscan}), which groups the reduced embeddings. It is an extension of DBSCAN \cite{ester1996density} which hierarchically groups clusters by different densities. It uses a soft-clustering approach, allowing noise to be modeled as outliers. 

Finally, topics are extracted with class-based TF-IDF (c-TF-IDF). The idea behind the original TF-IDF is to compare the importance of words between documents by computing the frequency of a word in a given document and also how prevalent the word is given all other documents. c-TF-IDF proposes to treat all documents in a single class as one single document. Important words are computed by measuring how frequent a word is in a class and how prevalent the word is in all different classes. The result would be an importance score for words within a class instead of a document. If we extract the most important words in a cluster, they are the description of a topic.

\end{document}